\documentclass[mathleft
]{an}
\usepackage{graphicx}
\usepackage{times}
\usepackage{amsmath}
\begin{document}

\Pagespan{952}{}
\Yearpublication{2013}%
\Yearsubmission{2005}%
\Month{11}%
\Volume{334}%
\Issue{9}%

\newcommand{\Babs}{\langle|B|\rangle}

\newcommand{\PSF}{\mathrm{PSF}}
\newcommand{\MTF}{\mathrm{MTF}}
\newcommand{\revone}{}

\title{Turbulent magnetic energy spectrum 
  and the cancellation function of solar photospheric magnetic fields}

\author{G. Marschalk\'{o}\inst{1,2}\fnmsep\thanks{Corresponding author:
  \email{G.Marschalko@astro.elte.hu}\newline}
\and  K. Petrovay\inst{1}
}
\titlerunning{Solar cancellation function}
\authorrunning{G. Marschalk\'{o} \& K. Petrovay}
\institute{
E\"{o}tv\"{o}s University, Department of Astronomy, Budapest, P.O. Box 32, 
H-1518, Hungary
\and 
Konkoly Observatory of the Hungarian Academy of Sciences
}



\abstract{%
A simple analytical relation of form $\alpha=2\kappa-1$ between
the magnetic energy spectral exponent $\alpha$ of the turbulent
magnetic field in the solar photosphere and its magnetic flux
cancellation exponent $\kappa$, valid under certain restrictive
assumptions, is tested and extended outside its range of validity in
a series of Monte Carlo simulations. In these numerical tests
artificial ``magnetograms'' are constructed in 1D and 2D by
superposing a discrete set of Fourier modes of the magnetic field
distribution with amplitudes following a power law spectrum and
measuring the cancellation function on these simulated magnetograms.
Our results confirm the validity of the analytical relation and extend
it to the domain $\alpha<-1$ where $\kappa\rightarrow 0$ as
$\alpha\rightarrow -\infty$. The observationally derived upper limit
of 0.38 on $\kappa$ implies $\alpha<-0.24$ in the granular size range,
apparently at odds with a small scale dynamo driven in the inertial
range.
}

\maketitle

\section{Introduction}

The structure of the magnetic field in the solar photosphere is still
far from being fully understood (for reviews see de Wijn et al. 2007,
Mart\'{\i}nez Pillet 2012). The two main methods of studying solar
magnetism are based on the Zeeman and Hanle effects, respectively. As
the above mentioned effects display different sensitivity to various
field configurations depending on their amplitude and spatial
organization,  we have a dichotomic view of the magnetism of the solar
photosphere. 

The basic properties of the magnetic network, consisting of kilogauss
strength flux tubes, have been clarified by longitudinal Zeeman
magnetometry  decades ago. The contribution of these elements to the
large scale unsigned flux density  $\Babs$ typically does not exceed a
few Gauss in quiet sun regions. Yet the analysis of the Hanle effect
depolarization of spectral lines indicates that the total value of
$\Babs$ is well in excess of 100\,G (Trujillo Bueno et al. 2004). Most
of this flux was, then, previously hidden to traditional Zeeman
magnetometry, presumably due to its fine scale turbulent structuring
that leads to the cancellation of the net Zeeman polarization signal
inside a resolution element. 

It has long been suspected that the sporadic internetwork (IN)
magnetic flux concentrations seen in longitudinal magnetograms
represent the observable part of this hidden or turbulent magnetic
field. Recent improvement in the resolution and sensitivity of
polarimeters, in particular the SDO/HMI {\revone (Solar Dynamics Observatory/Helioseismic and Magnetic Imager)}, 
Hinode/SP {\revone (Spectropolarimeter)} and Sunrise/IMaX {\revone (Imaging Magnetograph eXperiment)}
instruments, have led to a breakthrough in the 3D vector polarimetry
of the photospheric magnetic field, clearly demonstrating the presence
of a large number of IN flux concentrations. In contrast to the nearly
vertical, kG network flux tubes, these IN concentrations only reach
hectogauss field strengths and their orientation may be either more or
less horizontal or vertical.  Statistically, the distribution of
magnetic field orientations in this turbulent field seems to be more
or less isotropic, perhaps with some (currently debated) preference
towards the horizontal direction. 

Given that this observed IN field now represents a non-negligible
fraction of the formerly ``hidden'' turbulent flux, its detection
offers a unique chance to empirically study MHD turbulence in a
compressible, stratified plasma. In this respect, recent studies of
the {\it cancellation function}{\revone ,} $\chi(l_0)${\revone ,} of the photospheric
magnetic field are of interest. {\revone The function} $\chi(l_0)$ is defined as the 
unsigned flux density $\Babs$ seen at a finite resolution $l_0$ in a
longitudinal magnetogram, normalized to the intrinsic total unsigned
flux density $\Babs_0$: 
\begin{equation}
\chi(l_0)=\Babs(l_0)/\Babs_0. 
\end{equation}
\noindent 
Analyzing a
Hinode{\revone /}SP magnetogram, Pietarila Graham, Danilovic and Sch\"ussler
(2009) found that the shape of the cancellation function is a power
law $\chi(l_0)\propto l_0^{-\kappa}$ in the range 0.2 to 20 Mm, with
$\kappa=0.26$. The analysis was repeated by Stenflo (2011) upon
recalibrating the magnetogram, yielding a value $\kappa=0.127$,
independently confirmed also by the analysis of Hinode{\revone /}NFI
magnetograms. Recently, Pietarila and Pietarila Graham (2012) have
made an extensive comparative analysis of the cancellation functions
resulting from SoHO/MDI, SDO/HMI and Hinode/SP magnetograms, finding
that the derived $\kappa$ values are heavily influenced by instrument
noise, seeing effects, net flux imbalance in the field and by the
proper exclusion of field components not belonging to the turbulent
field (such as netwok elements). They find that the $\kappa$ values
derived for quiet regions decrease with improving instrument quality,
i.e. along the MDI $\rightarrow$ HMI $\rightarrow$ SP series, so that
the SP value of 0.38 can be considered a conservative upper bound for
the true value of the cancellation exponent. Lower values reported in
earlier studies were apparently due to improper masking of network
fields and/or the use of fields of view with a higher flux imbalance
(i.e. less typical quiet sun areas).

As the most important theoretical tool in the study of the scaling
behaviour of turbulent flows is the energy spectral function, the
question naturally arises how $\chi({\revone l_0})$ is related to the magnetic
energy spectral function, written as  $E_k\sim k^\alpha$?

In Section 2 we derive a simple analytical relation between the
cancellation exponent{\revone ,} $\kappa${\revone ,} and the spectral exponent{\revone ,} $\alpha$,
valid under some assumptions. The relation is extended beyond the
limit of validity of the analytic formula in a set of Monte Carlo
simulations in Section 3. Conclusions are drawn in Section 4.

\begin{figure}[!t]
\includegraphics[width=\columnwidth,height=50mm]{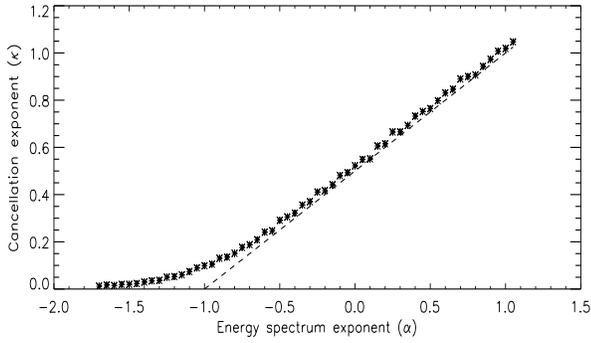}
\caption{Cancellation exponent $\kappa$ as a function of the magnetic
energy spectral exponent $\alpha$ in the 1D case, with the analitically
derived relation (dashed line). }
\label{label1}
\end{figure}

\section{Analytical treatment of the problem}

Consider a fluctuating 2D field of zero mean, such as the {\it
residual} magnetic field upon subtracting the large scale mean field
from a magnetogram.  If the actual longitudinal residual magnetic
field distribution is denoted by $B_0(x,y)$ then the residual
magnetogram observed at a finite resolution $l_0$ is
\begin{equation}
\label{eq_bxy}
B(x,y) = \iint B_0(x',y')\, \PSF(x-x',y-y')\, dx'\, dy'
\end{equation}
where $\PSF$ is the point spread function for the given resolution $l_0$
{\revone , and x,y are Cartesian coordinates of the magnetogram}.
The Fourier transform of {\revone equation (\ref{eq_bxy})} is
\begin{equation}
\hat{B}(k_x,k_y) = \hat{B_0}(k_x,k_y)\cdot \MTF(k_x,k_y; k_0)
\end{equation}
where the modulation transfer function $\MTF$ is the Fourier transform
of the $\PSF$, {\revone $k_x^2 + k_y^2 = k^2 $} and $k_0=\pi/l_0$. For
statistical isotropy in the plane the $\MTF$ only depends on the
modulus of the wave number ($k$).
{\revone  For simplicity we also assume that there is only one
characteristic resolution scale involved: in this case the MTF will
only depend on the combination $k/k_0$.}

Let the spectrum of the magnetic energy (in the line of sight
component) as a function of wave number be $E_k$. 
{\revone From elementary turbulence theory we have}
\begin{equation}
	E_k = {1\over2}\,\hat{B}(k)\, \hat{B}^{*}(k)
\end{equation}
(As with $B$, $E_k$ is the spectrum observed at finite resolution while
$E_{0,k}$ is the actual spectrum{\revone \ and $\hat{B}^{*}(k)$ is the 
complex conjugate of $\hat{B}(k)$.})

Assuming a power law turbulent magnetic energy spectrum
\begin{equation}
	E_{0,k} = E_{0,k_0}\, \left({k \over k_0}\right)^{\alpha}
\end{equation}
with $\alpha>-1$,{\revone the total magnetic energy observed at resolution
$l_0$ can be calculated by means of the $\MTF$ in the following way:

\begin{eqnarray}
	E(k_0) = \int_{0}^{\infty}E_k\, dk = \int_{0}^{\infty}E_{0,k}\, 
	|\MTF(k/k_0)|^2\, dk \nonumber \\
	= k_0 E_{0,k_0} \int_{0}^{\infty}\, {\tilde k}^{\alpha}\, 
	|\MTF({\tilde k})|^2\, d{\tilde k}
	\, \text{    } 
	\label{eq:2ints} 
\end{eqnarray}
where we used the substitution $\tilde k=k/k_0$.

The derivative of equation (\ref{eq:2ints}) by $k_0$ is:
\begin{equation}
	\label{eq:Ederiv}
	E'(k_0) = E_{0,k_0} \int_{0}^{\infty}\, {\tilde k}^{\alpha}\, 
	|\MTF({\tilde k})|^2\, d{\tilde k}
\end{equation}
This demonstrates that, independently of the form of the MTF,  the
rate of increase in magnetic energy seen with increasing resolution is
proportionate to the value $E_{0,k_0}$ of the magnetic energy spectral
function at the inverse resolution, $k_0$.}

To find a relation between this {\revone rate of increase} and the
cancellation function  we need to make a further assumption regarding
the self-similarity (fractal nature) of the {\revone magnetic field
distribution} $B_0(x,y)$.  {\revone Let us introduce the notation 
$F=\langle B^2 \rangle / \langle |B| \rangle^2$. 
The probability density function (PDF)} of $|B|$
will determine the value of $F$, but for a scale invariant (fractal)
field distribution $F$ should be resolution independent. This suggests
that $\langle B^2 \rangle \sim k_0^{2\kappa}\, \mathrm{if}\, \langle
|B| \rangle \sim k_0^{\kappa}$. The derivative of this yields by
Eq.\,(\ref{eq:Ederiv}) $E_{0,k} \sim k^{2 \kappa -1}$ i. e.
\begin{equation}
\alpha = 2\, \kappa - 1.  \label{eq:analrel}
\end{equation}

Note that this formula is the non-intermittent special case of the
more general relationship derived by Vainshtein et al. (1994). While
the  derivation given above is more restricted in its assumptions
regarding intermittency, it does include the effects of finite
instrumental resolution and its transparency makes it easy to see the
limitations of validity of equation (\ref{eq:analrel}). Indeed, it
should be stressed that this relationship is valid only for
$\alpha>-1$. For $\alpha<-1$ equation (\ref{eq:analrel}) would result
in negative values for $\kappa$ which is clearly impossible in view of
the meaning of the cancellation function. Indeed, for $\alpha<-1$ the
integral on the r.h.s. of equation {\revone (\ref{eq:Ederiv})
would diverge at its lower limit (as $|MTF|(k\rightarrow 0)\rightarrow
1$).} In this case it needs to be taken into account that the power
law range in the spectrum can  only extend down to an integral scale
$k_1$, replacing zero as the lower bound of the integral, and the
value of the integral will be dominated by the contribution near
$k_1$, asymptotically making $E(k_0)$ and $\chi$ independent of $k_0$.
Our expectation, then, is that $\kappa\rightarrow 0$ as
$\alpha\rightarrow -\infty$. Due to the dependence on a preferred
scale $1/k_1$ the cancellation function is expected to deviate from
power law behaviour in this regime.

We further note that, {\revone depending on the form of the MTF and
the value of $\alpha$, the integral in equation (\ref{eq:Ederiv}) may
also diverge at its upper limit.  This may set an upper limit on
$\alpha$ for the validity of our analytical formula (and of the power
law form of $\chi$). As, however, this depends on the particular form
of the MTF used, this upper limit is less generic than the lower bound
$\alpha>-1$ discussed above.}

\begin{figure}[!t]
\includegraphics[width=\columnwidth,height=50mm]{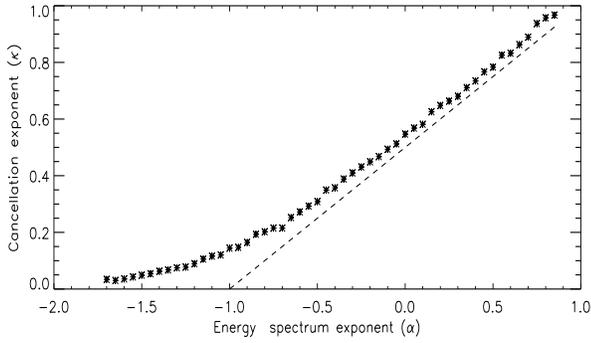}
\caption{Cancellation exponent $\kappa$ as a function of the magnetic
energy spectral exponent $\alpha$ in the 2D case, with the
analitically derived relation (dashed line). }
\label{label2}
\end{figure}

\section{Monte Carlo simulations}

To demonstrate the validity of the analytical relation
(\ref{eq:analrel}) as well as to extend the $\kappa$--$\alpha$
relationship into the range $\alpha\la -1$ we set up a series of Monte
Carlo simulations. First we do this for the 1D case, for conceptual
simplicity and numerical advantages.

We represent the magnetic field by a discrete set of $n$ modes,
equidistant with spacing 1 on a base-2 logarithmic scale. Assuming
power law energy spectra with different exponents, the energy between
$k_0$ and $2\, k_0$ is
\begin{equation}
	\int_{k_0}^{2\, k_0}E_k\, dk = \int_{\log_2 (k_0)}^{\log_2 (k_0) + 1} 
	k E_k\,  d\log_2 k
\end{equation}

Since $k\, E_k\sim k^{1+\alpha}$ we have $B_k \sim k^{(1+\alpha)/2}$,
so the Fourier amplitude of each mode is: 
\begin{equation} 
\hat B_j = \hat B_{j,0}\, e^{i\, \phi_j}\, \text{ where }\,  
\hat B_{j,0} \sim k^{(1+\alpha)/2}
\end{equation} 
where $j=1..n$ and the phases $\phi_j$ were assumed to be random.
(We use $n=14$ in the 1D case.)
The computation was performed on a grids of $N=2^n$ points, the modes
considered having $k_j = 2\pi/2^j\, (j=1..n)$. $B$ reads
\begin{equation}
	B = \sum_{j = 1}^{n} \hat B_j\, e^{ik_j x} + 
	\sum_{j = 1}^{n} \hat B_j^\ast\, e^{-ik_j x}
\end{equation}

Having generated the function B on the full grid, we smear it with
various ``masks'' or grids of coarser resolution ($2^j$ grid points
with $j=1..n$), at different random positions, and we calculate the
resulting flux density $\Babs$. The $\alpha$ values are taken from the
range [-3,3], with a denser sampling in the interval [-0.9,0]. For
each value of $\alpha$, the computation is done for a large number of
random phase sets $\{\phi_j\}$ and mask positions, averaging the
results. The exponent of the cancellation function was determined by a
power law fit to all points except the two largest scales.

In the next step the calculation is performed again for the 2D case
with the same method. Due to the higher computational requirements the
maximum grid size, and hence the number of modes was lower ($n = 12$)
than in the 1D case.

\begin{figure}[!t]
\includegraphics[width=\columnwidth,height=50mm]{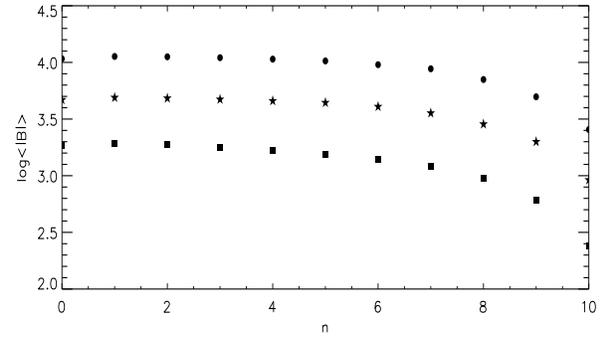}
\caption{Cancellation function with different turbulent magnetic
energy spectra. Dots: $\alpha = -1.7$; stars: $\alpha = -1.6$;
squares: $\alpha = -1.5$.}
\label{label3}
\end{figure}

The resulting $\alpha$--$\kappa$ relation is plotted in Figures 1 and
2 for the 1D and 2D case, respectively. It is apparent that within its
range of validity the analytical relation (\ref{eq:analrel}) indeed
provides an excellent description of the relationship, down to
$\alpha$ values of $-0.7$ or so. Towards large negative values of the
spectral exponent we have $\kappa\rightarrow 0$ as expected.

Figures 3 and 4 present the cancellation functions obtained with
different $\alpha$ values. It is apparent that for values within the
scope of the analytical relation (Figure 4) the cancellation function
can be fairly well estimated by a power law. On the other hand, for
lower $\alpha<-1$ (Figure 3) the cancellation function shows a
significant deviation from power law behaviour, as expected (cf. the
discussion at the end of Section 2.)

The conservative upper limit of $\kappa<0.38$ obtained by Pietarila
and Pietarila Graham (2012) implies $\alpha<-0.24$ according to our
analytical relation (\ref{eq:analrel}), confirmed by the numerical
results. The much more stringent result obtained by Stenflo (2011), on
the other hand, would imply $\alpha\simeq -0.95$ based on our
numerical results. (The analytical relation would give $-0.746$).

\section{Conclusion}

Under the assumption of a self-similar (fractal) magnetic field
structure we have derived a simple analytical relationship, equation
(\ref{eq:analrel}), between $\kappa$ and $\alpha$. The relation is
expected to be valid for $\alpha>-1$. The relation is tested and
extended outside its range of validity in a  series of Monte Carlo
simulations. In these numerical tests artificial ``magnetograms'' are
constructed in 1D and 2D by superposing a discrete set of Fourier
modes of the magnetic field distribution with amplitudes following a
power law spectrum and measuring the cancellation function on these
simulated magnetograms. Our results confirm the validity of the
analytical relation and extend it to the domain $\alpha<-1$ where
$\kappa\rightarrow 0$ with decreasing $\alpha$ values.

The observationally derived upper limit of 0.38 on $\kappa$ implies
$\alpha<-0.24$. The lowest $\kappa$ values detected in any magnetogram
field to date would correspond to $\alpha\simeq -0.95$. These findings
provide evidence that the magnetic energy spectral function is a
decreasing function of wavenumber in the granular size range (0.2 to
20 Mm), in contrast to the prediction of small scale dynamo
simulations where $\alpha>0$ is found (Schekochihin et al. 2007,
Pietarila Graham et al. 2010). The limits we derived do not exclude
the possibility of an $\alpha=-1$ spectrum which is the hallmark of a
larger scale field being passively advected by turbulent motions. This
might indicate that the photospheric turbulent magnetic field results
from passive field amplification; however, in this case we would
expect $\Babs\sim|\langle B\rangle|$ which is not observed (Lites
2011). This suggests that the turbulent magnetic field in the solar
photosphere does originate in a dynamo but this dynamo does not
operate in the inertial range as in current simulations but rather at
or above the integral scale. This may be due to the fact the magnetic
Prandtl number $P_m$ in the solar plasma is much lower than in current
numerical simulations. There are indications that at such low values
of{\revone ,} $P_m${\revone ,} the critical magnetic Reynolds number{\revone ,} $R_m${\revone ,} for dynamo action
is much higher than for $P_m\simeq 1${\revone ,} which may impede the operation
of an inertial range dynamo. At larger scales, however, $R_m$ may be
high enough to drive a dynamo; or alternatively a small scale dynamo
driven by a fundamentally different process might be at work at or
above the integral scale.

Our constraints on $\alpha$ are in accordance with the value of
$\alpha\simeq -1.3$ derived by Abramenko et al. (2001) for quiet sun
regions from a direct analysis of magnetograms.

The single most important remaining restriction in our numerical
results is the assumption of random phases {\revone (i.e. lack of
intermittency).} The photospheric turbulent magnetic field is known to
be distributed in a very distinctive pattern, forming IN field
concentrations, following intergranular lanes and mesogranular
structure. This is very different from the amorphous superposition of
Fourier modes with random phases. Whether, and to what extent
non-random phases might influence our inferences should be the subject
of future research.

\begin{figure}[!t]
\includegraphics[width=\columnwidth,height=50mm]{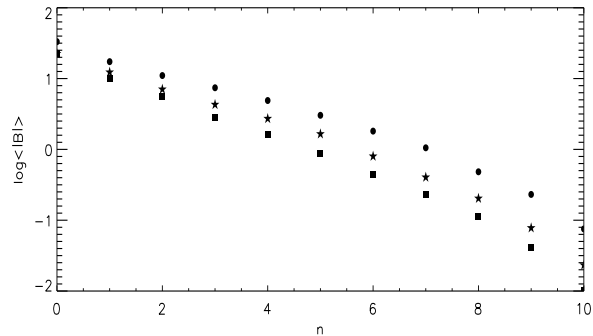}
\caption{Cancellation function with different turbulent magnetic
energy spectra. Dots: $\alpha = -0.7$; stars: $\alpha = -0.6$;
squares: $\alpha = -0.5$.}
\label{label4}
\end{figure}

\acknowledgements
This research was supported by the Hungarian Science Research Fund
(OTKA)  grants no.\ K81421 and K83133. This project has been also
supported by ESTEC Contract No. 4000106398/12/NL/KML.

\end{document}